\documentclass[aps,pre,showkeys,superscriptaddress,floatfix,reprint]{revtex4-1}
\usepackage{graphicx}
\usepackage{epsfig}
\usepackage{amsmath}
\usepackage{graphics}
\usepackage{latexsym}
\usepackage{color}
\usepackage[update,prepend]{epstopdf}

\newcommand{\la}{\left\langle}
\newcommand{\ra}{\right\rangle}
\begin{document}
\title{The smectic phase in semiflexible polymer materials: \\
A large scale Molecular Dynamics study}
\author{Andrey Milchev}
\affiliation{Institute for Physical Chemistry, Bulgarian Academia of Sciences, 1113, Sofia, Bulgaria}
\affiliation{Institute of Physics, Johannes Gutenberg University Mainz, Staudingerweg 7, 55128 Mainz, Germany}

\author{Arash Nikoubashman}
\affiliation{Institute of Physics, Johannes Gutenberg University Mainz, Staudingerweg 7, 55128 Mainz, Germany}

\author{Kurt Binder}
\affiliation{Institute of Physics, Johannes Gutenberg University Mainz, Staudingerweg 7, 55128 Mainz, Germany}

\begin{abstract}
Semiflexible polymers in concentrated lyotropic solution are studied within a 
bead-spring model by molecular dynamics simulations, focusing on the emergence
of a smectic A phase and its properties. We systematically vary the density of 
the monomeric units for several contour lengths that are taken smaller than the
chain persistence length. The difficulties concerning the equilibration of such 
systems and the choice of appropriate ensemble (constant volume versus constant 
pressure, where all three linear dimensions of the simulation box can fluctuate 
independently) are carefully discussed. Using HOOMD-blue on graphics processing 
units, systems containing more than a million monomeric units are accessible, 
making it possible to distinguish the order of the phase transitions that occur.
While in this model the nematic-smectic transition is continuous, the transition
from the smectic phase to a related crystalline structure with true
three-dimensional long-range order is clearly of first order. Further, both
orientational and positional correlations of monomeric units are studied as well
as the order parameters characterizing the nematic, smectic A, and crystalline
phases. The analogy between smectic order and one-dimensional harmonic crystals
with respect to the behavior of the structure factor is also explored. Finally,
the results are put in perspective with pertinent theoretical predictions and
possible experiments.
\end{abstract}

\maketitle

\section{Introduction}
Liquid-crystalline phases of semiflexible polymers are materials with great 
potential for various applications \cite{Krigbaum, Ciferri, Donald}. The 
chemical structure of these materials is typically rather complicated and a 
detailed understanding of structure-properties relations is often rather 
incomplete. While for liquid crystals formed from small molecules a chemically 
realistic atomistic molecular modeling has become possible \cite{Glaser, Palermo,
Sidky}, the large length scales involved in semiflexible polymers require the
use of coarse-grained models \cite{Voth}. Typical coarse-grained models of
semiflexible polymers consist of $N$ ``effective monomeric units'' with diameter
$\sigma$ and distance $\ell_{\rm b}$, evenly placed along the contour of the
chain. In such a representation, the contour length, $L$, of the macromolecule
is given by $L = (N-1) \ell_{\rm b}$. The chain stiffness, which is responsible
for the emergence of liquid crystallinity, is described by the persistence length
$\ell_{\rm p}$, which is much larger than the length $\ell_{\rm b}$ of the
effective bonds, and can be of the same order as $L$. Previous simulations using
such coarse-grained models have been rather successful in the description of
the isotropic-nematic transition in lyotropic solutions (assuming an implicit
description of the solvent) \cite{SEAMKB, SEAMPVKB, AMSEKBAN, popadic:sm:2018,
popadic:arx:2018}.

For lyotropic solutions of rigid rods ($\ell_{\rm p}/L = \infty$), a subsequent
transition to the smectic-A phase has been identified with increasing density
in both simulations \cite{frenkel:nat:1988} and in experiments \cite{Wen}. It is
possible that smectic phases occur as well in concentrated solutions of less
stiff semiflexible polymers ($\ell_{\rm p} \gtrsim L$), but the conditions
necessary to find such phases are not understood in sufficient detail
\cite{Tkach1, Tkach2, Tkach3}. Simulations so far considered short chains of
strongly overlapping beads ($\ell_{\rm b} \ll \sigma$) in order to model
slightly flexible rod-like molecules \cite{Cinacci, Schoot, Schoot3}. While in
those studies smectic-A phases were found, the exploration of a bead-spring
model of semiflexible chains with $\ell_{\rm b} \approx \sigma$ indicated the
emergence of smectic-C order, at least in the two-dimensional case \cite{AMSEKB,
AMKB, KBSEAM}. Distorted forms of smectic order were also found under spherical
confinement, both in the bulk of the sphere \cite{Vega1,Vega2} and in thin
spherical shells \cite{milchev:polymer:2018, Khadilkar}. Recent simulations
\cite{AMSEKBAN} indicated the occurrence of smectic order in simulations in the
bulk at melt densities ($\rho \geq 0.7$) for chains with $\ell_{\rm p} \gg L$,
but a detailed study of the smectic phase for this model has not yet been
performed.
 
Such a study is a challenge for molecular dynamics (MD) simulations since the
number of smectic layers, $n$, has to be much larger than unity in order to avoid
finite-size effects, and the wave length $\Lambda$, characterizing the layered
smectic structure, is of the same order as $L$. If the layering occurs in
$z$-direction, the simulation box linear dimension $L_z$ should be strictly 
commensurate with $n\Lambda$. However, the precise value of $\Lambda$ is not known 
beforehand, and also the simulation setup must not suppress statistical
fluctuations of $\Lambda$ which are an important ingredient of the problem. As a 
consequence, one should not use the constant volume (${\cal N}VT$) ensemble 
(${\cal N}$ is the total number of chains, $V$ is the system volume, and $T$ is
the absolute temperature) where the linear dimensions $L_x$, $L_y$, $L_z$ of the
simulation box are held fixed. On the other hand, if one uses the constant pressure
(${\cal N}PT$) ensemble, one must ensure that also for $L_z \gg L_x, L_y$ the
transverse linear dimensions $L_x$ and $L_y$ are sufficiently large to avoid
instabilities of the algorithm and possible distortion of the ordering.

From these considerations it is clear that such simulations require the use of 
an efficient but also versatile simulation software allowing the study of systems
containing of the order of million effective monomeric units.
Previous efforts using a model with strongly overlapping beads, restricted
attention to a single chain length ($N=9$) using ${\cal N}=600$ chains in total
\cite{Cinacci}, or ${\cal N}=4464$ chains with $N=17$ beads \cite{Schoot}, or
${\cal N}=4608$ chains with $13 \le N \le 21$ \cite{Schoot3}. The aim of the
present work, however, is the study of much larger systems, e.g., with up to
${\cal N}N=1694784$ monomeric units to ensure that the results are not affected
by systematic finite size effects. Our work did become feasible owing to the
availability of the HOOMD-blue software package \cite{Anderson,Glotzer}.

The remainder of this manuscript is organized as follows. In Sec.~\ref{sec:model}
we shall summarize the methodology to equilibrate the model system and to
characterize its liquid-crystalline order. Section~\ref{sec:results} then describes
the  results while Sec.~\ref{sec:conclusions} contains our conclusions.


\section{Model and methods}
\label{sec:model}
Our model choice is dictated by the fact that semiflexible polymers in lyotropic
solutions may exhibit vastly different conformations in the various phases that
are expected to occur. For small enough polymer concentration in the solution,
an isotropic phase occurs where the end-to-end vectors of the chains are randomly
oriented. For $\ell_{\rm p} \ll L$, these chains have coil-like conformations,
whereas in the inverse limit $\ell_{\rm p} \gg L$, they rather resemble flexible
rods. In the nematic phase, the chains are always stretched out strongly and have
a root mean square end-to-end distance $\la R_{\rm e}^2 \ra^{1/2}$ not much
smaller than $L$ (we disregard here the occurrence of ``hairpin`` conformations
that are found for $\ell_{\rm p} \ll L$ in the nematic phase close to the
isotropic-nematic transition \cite{AMSEKBAN,Vroege}). Being interested in smectic
phases with periodicity $\Lambda \sim L$, we focus on the choices $N=8$, $12$, 
and $16$ here. We choose the same model as in our previous work \cite{SEAMKB,
SEAMPVKB, AMSEKBAN}, so that properties of individual chains, etc., in the various
phases can be meaningfully compared.  

We use here the Kremer-Grest model \cite{Grest,K_G} extended by a bending
potential to control chain stiffness. The interaction between any pair of beads 
is purely repulsive and of short range,
\begin{equation}
	U_{\rm WCA}(r) = 4\epsilon \left[\left(\frac{\sigma}{r}\right)^{12} -
	\left(\frac{\sigma}{r} \right)^6 + \frac{1}{4} \right],
	r \le r_c,
	\label{eq_1}
\end{equation}
where $r$ is the distance between a pair of beads, and $r_{\rm c} \equiv
2^{1/6}\,\sigma$ is the cutoff distance of the potential ($U_{\rm WCA}(r >
r_{\rm c}) = 0$). The parameter $\epsilon$ controls the strength of the potential,
and it is chosen as our unit of energy. The bead diameter, $\sigma$, is chosen as
the unit of length, and the bead mass, $m$, as the unit of mass.

Further, neighboring beads along a chain interact through the finitely-extensible 
nonlinear elastic (FENE) potential \cite{Grest, K_G}, 
\begin{equation}
	U_{\rm FENE}(r) = -0.5 k r_0^2 \ln \left[1 - \left(\frac{r}{r_0}\right)^2   
	\right], r < r_0,
	\label{eq_2}
\end{equation}
with spring constant $k = 30\,\epsilon/\sigma^2$. The parameter $r_0 = 1.5\,\sigma$
controls the maximum extension of the spring, and $U_{\rm FENE} (r>r_0) = \infty$.
The distance $\ell_{\rm b}$ between two consecutive beads is $\ell_{\rm b} \approx
0.97\,\sigma$ for the chosen model parameters (the precise value depends slightly
on density, temperature and chain stiffness). 

The bending potential depends on the angle $\theta_{ijk}$ formed between two 
consecutive bond vectors $\mathbf{a}_i = \mathbf{r}_j - \mathbf{r}_i$ and
$\mathbf{a}_j = \mathbf{r}_k - \mathbf{r}_j$ with $j = i+1$ and $k=j+1$ as
\begin{equation} 
	U_{\rm bend}(\theta_{ijk}) = \kappa [1 - \cos(\theta_{ijk})],
 	\label{eq_3} 
\end{equation}
where an angle of $\theta_{ijk}=0^\circ$ corresponds to three beads in a line.
The strength of this potential $\kappa$ is chosen in the range $\kappa =
8\,\epsilon$ to $\kappa = 128\,\epsilon$.

The persistence length of the polymers, $\ell_{\rm p}$, is defined in terms of
$\la \cos(\theta_{ijk})\ra$ \cite{Hsu} 
\begin{equation}
	\ell_{\rm p} = -\frac{\ell_{\rm b}}{\ln\la\cos\theta_{ijk}\ra} .
	\label{eq_4}
\end{equation}
One can show for large $\kappa$ and dilute solutions, where chain
interactions can be neglected, that $\ell_{\rm p} \approx \ell_{\rm b} \kappa/
(k_{\rm B}T)$ (with Boltzmann's constant $k_{\rm B}$ and temperature $T$).
In concentrated solutions or melts with liquid crystalline order, however, the
actual persistence length found from Eq.~(\ref{eq_4}) can be significantly
enhanced in comparison with this estimate for the ``bare'' persistence length
\cite{AMSEKBAN}.

This model has been studied by MD simulations \cite{Tildesley,Rapaport}. Both
${\cal N}VT$ and ${\cal N}PT$ ensembles have been used, employing a time step
$\Delta t = 0.002\,t_{\rm MD}$, with intrinsic time unit of MD, $t_{\rm MD} =
\sqrt{m\sigma^2/\epsilon}$. In the ${\cal N}VT$ simulations, temperature was
controlled through the standard Langevin thermostat \cite{Grest, K_G} as in our
previous work \cite{SEAMKB, SEAMPVKB, AMSEKBAN}. For the ${\cal N}PT$ simulations
we use a Martyna-Tuckerman-Tobias-Klein barostat \cite{Martyna,Klein}, where the
equations of motion are time reversible and leave the phase space measure
invariant. The coupling constants for the thermostat and barostat were chosen
as $t_T = 0.5$ and $t_P = 1.0$, respectively. As emphasized already in
the introduction, because of the necessity of very large system sizes, this work
becomes possible only due to the availability of HOOMD-blue \cite{Anderson,Glotzer}.

The ${\cal N}PT$ ensemble was chosen in a variant where all linear dimensions of
the box were identical, so a cubic $L_x \times L_y \times L_z$ shape of the
simulation box was enforced, as well as in a variant where the linear dimensions
$L_x$, $L_y$, and $L_z$ were allowed to fluctuate independently, employing hence
a box of rectangular slab shape. Periodic boundary conditions in all directions
were used throughout. The choice of a cubic box in the ${\cal N}VT$ ensemble is 
appropriate for the system in its isotropic phase, but becomes questionable in 
the liquid-crystalline phases. This is explicitly seen when we record the 
pressure tensor $P^{\alpha \beta}$ and its fluctuations ($\alpha, \beta = x, 
y, z$). The instantaneous pressure is found from the Virial expression 
\cite{Tildesley}
\begin{equation}
	P^{\alpha,\beta} = \rho k_{\rm B}T \delta^{\alpha, \beta} + \frac{1}{3V}
	\sum_i r_i^\alpha F_{\rm tot}^\beta (\mathbf{r}_i)
	\label{eq_5}
\end{equation}
where the sum extends over all beads in the system, the density $\rho$ is given 
by $\rho \equiv N{\cal N}/V$, and $\mathbf{F}_{\rm tot}(\mathbf{r}_i)$ is the
total force acting on bead $i$ at position $\mathbf{r}_i$ due to the potentials
Eqs.(\ref{eq_1})-(\ref{eq_3}).

\begin{figure*}[htb]
	\includegraphics[width=16.5cm]{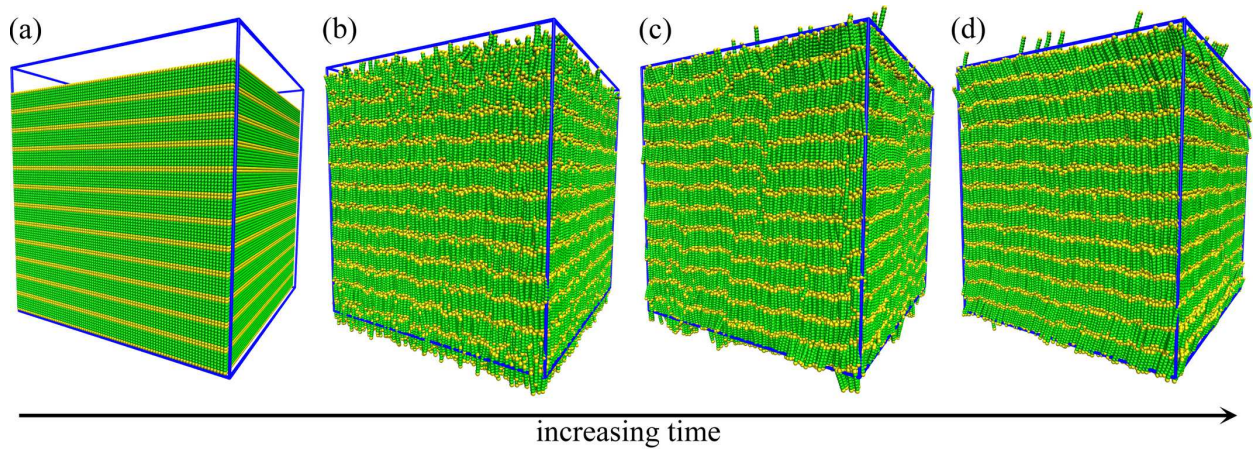}
	\caption{Formation of an ordered phase of semiflexible polymers with $N=8$, 
	$\kappa = 96$, ${\cal N}=101568$ in a $98.06 \times 98.06 \times 104$ box, with
	$\rho=0.8125$ at $k_{\rm B}T=0.5$. The initial state (a) is a 
	regular arrangement of 12 layers of straight chains with some extra volume on 
	top. At $t=100$ (b), this density inhomogeneity is removed but only local disorder
	in the $n=12$ layers has appeared. At $t = 1400$ (c), however, there is a local
	defect in the layering, a new $13^{th}$ layer starts to form and a stable 13
	layers persist as a final structure. The last snapshot refers to $t=20000$ (d).}
	\label{fig_1}
\end{figure*}

We initialize the chains as straight rods along the $z$-axis, and then arrange
the chains on a square lattice (or, alternatively, on a triangular lattice). The
lattice spacing is chosen such that the desired density $\rho$ is reached, and
$n$ of these layers of stretched chains are then put on top of each other. The
choice of the proper value of $n$ (or $L_z$) is not obvious a priori, because
the precise value of the smectic period, $\Lambda$, that eventually develops is
not known in beforehand. It is necessary to make sure that a slightly incorrect
choice of $n$ does not prevent the approach towards the correct equilibrium
state. For example, the choice $N=8$, ${\cal N} = 101568$ and $L_x=L_y=L_z=100$,
leads to a density of $\rho = 0.8125$. If one tries to equilibrate such a system,
initialized with a regular arrangement comprising $n=12$ smectic layers, one
obtains a strongly distorted smectic structure. Choosing $L_z$ slightly larger,
namely $L_z=104$, then $13$ layers would  fit better than only $12$ layers, and
indeed we observe a transition from $n=12$ to $n=13$ during the run,
Fig.~\ref{fig_1}.

The system shown in Fig.~\ref{fig_1} has been initialized using a square lattice
arrangement of chains in the $xy$-plane, stretched out along the $z$-axis. This
configuration is clearly not similar to the crystalline ground state of our model,
since a regular packing of rigid rods would rather result in a triangular lattice
structure in the cross-sectional $xy$-plane. In order to test for a possible bias
in our results, due to the choice of the initial state, we have also carried out
runs with a triangular lattice arrangement of rod-like polymers (and choosing
then $L_x/L_y = 2/\sqrt{3}$ so that the triangular lattice arrangement is
compatible with the periodic boundary condition). We have found that the memory
of the initial crystalline chain arrangement is quickly lost for the densities of
interest. The initial layering does not create an undue bias either. We have 
tested this fact by creating artificial states with hexagonal order in the $xy$-
plane but disorder in the $z$-coordinates of the center of mass positions of the
stretched out polymers. Hence, it appears that for the densities of interest the
resulting nematic or smectic structures are developing with the proper order
irrespective of this initial disorder. Alternatively, one could also attempt to
produce ordered phases starting from fully disordered isotropic chain configurations.
However, this task is quite challenging since the growth of ordered domains is a
rather slow process to be convenient for simulations. 

The time evolution in Fig.~\ref{fig_1} shows that the extra volume for $n=12$ 
layers in a $98 \times 98 \times 104$ box is rapidly filled at first, and then 
large-scale defects in the structure form by which the system is able to create 
an extra layer and form the more stable arrangement with $n=13$ layers. However,
the final snapshot clearly reveals that for the chosen conditions a small
mechanical deformation (``buckling'') is still present: this can be avoided only 
by using a ${\cal N}PT$ rather than ${\cal N}VT$ simulations.

\begin{figure}
	\includegraphics[width=7.5cm]{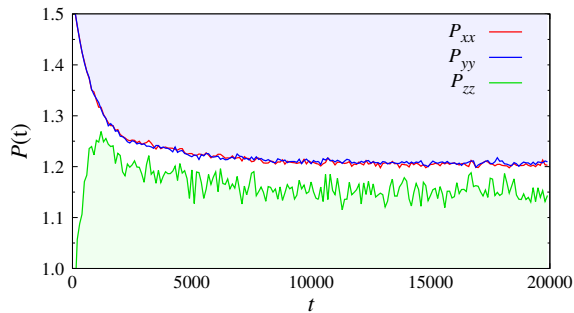}
	\caption{Time evolution of pressure tensor components parallel ($P_{zz}$) and 
	perpendicular ($P_{xx}, P_{yy}$) to the nematic director, for the system  
	$N=8$, ${\cal N}=101568$, $\kappa = 16$, $L_x=L_y=98.06$, $L_z=104$, and
	$k_{\rm B}T=1.0$.}
	\label{fig_2}
 \end{figure}

This conclusion is enforced by following the time evolution of the pressure 
tensor components, Fig.~\ref{fig_2}. It is seen that a time of order
$10^4\,t_{\rm MD}$ is necessary to relax the system until the individual pressure 
components become independent of time but a systematic difference $\delta P 
\equiv P_{zz} - (P_{xx}+P_{yy})/2$ of order $0.05$ remains when we work in 
the ${\cal N}VT$ ensemble.

\begin{table*}[htbp!]
\begin{tabular}{|c|c|c|c|c|c|c|c|c|c|c|c|}
\hline
$ L_x$ & $ L_y $ &  $L_z$ & $P_{xx}$ & $P_{yy}$ & $P_{zz}$ & $\rho$ & $S$ &
$\la R_{\rm e}^2\ra$ & $\la R_{{\rm e},z}^2\ra$ & $\la R_{{\rm e}, \perp}^2\ra$
& $\langle \ell_{\rm b}^2\rangle$ \\
\hline
134.06 & 134.06 & 134.06 & 1.285 & 1.285 & 1.269 & 0.680 & 0.913 & 203.5 &
197.5 & 5.95 & 0.938 \\
134.1 & 134.1 & 134.1 & 1.282 & 1.283 & 1.266 & 0.679 & 0.912 & 203.5 & 197.5 &
5.95 & 0.938 \\
135.7 & 135.6 & 131.0 & 1.280 & 1.283 & 1.282 & 0.679 & 0.912 & 203.4 & 197.5 &
5.95 & 0.938 \\
108.1 & 108.1 & 204.3 & 1.285 & 1.285 & 1.281 & 0.680 & 0.911 & 203.4 & 197.3 &
6.18 & 0.938 \\
\hline
\end{tabular}
\caption{Simulation results for $N=16$ and $\kappa = 64$. Linear dimensions of
the simulation box $L_x$, $L_y$, and $L_z$. Diagonal components of the pressure
tensor, $P_{xx}$, $P_{yy}$, and $P_{zz}$. Monomer density, $\rho$, nematic order
parameter, $S$, mean-square end-to-end distance, $\la R_{\rm e}^2\ra$, and its
$z$-component, $\la R_{{\rm e},z}^2\ra$, as well as its transverse component,
$\la R_{{\rm e}, \perp}^2\ra$. Mean-square bond length, $\la \ell_{\rm b}^2\ra$.}
\label{table_1}
\end{table*}

Already in the nematic phase such systematic differences start to show up as 
demonstrated in Table~\ref{table_1}, cf.  Ref.~\cite{AMSEKBAN}. The results are
for the case $N=16$, $\kappa = 64$ at $k_{\rm B}T = 1.0$. The first line uses
the ${\cal N}VT$ ensemble with a cubic box. Note that $P_{zz}$ is slightly smaller
than $P_{xx} \approx P_{yy}$. The second line shows ${\cal N}PT$ results for the
choice $P = 1.28$ but allowing only uniform volume fluctuations. The initial
conformation was taken here from the constant volume ensemble. The third line
shows also ${\cal N}PT$ results for $P = 1.28$, but now $L_x$, $L_y$, and $L_z$
can fluctuate independently of one another. Note that now the desired result
$P_{xx} \approx P_{yy} \approx P_{zz}$ holds within the statistical error (which
here is about $10^{-3}$). The final line shows ${\cal N}VT$ results for an
elongated box, linear dimensions taken from an equilibrated configuration in
the ${\cal N}PT$ ensemble (for the choice $P=1.28$). Clearly, the uniaxial
symmetry of the nematic order with the director along the $z$-axis is incompatible
with a cubic box, so independent fluctuations of $L_x$, $L_y$ and $L_z$ in the
${\cal N}PT$ ensemble are needed. Gratifyingly, in the nematic phase the results
for the nematic order parameter $S$ as well as the chain linear dimensions
$\la R_{{\rm e},z}^2\ra$ and $\la R_{{\rm e}, \bot}^2\ra$ in the ${\cal N}VT$
ensemble agree with their counterparts in the ${\cal N}PT$ ensemble. Therefore
it is useful to start with a study in the ${\cal N}VT$ ensemble (which is
computationally somewhat easier) to get a first orientation of the present problem.

\begin{figure*}[htb]
	\includegraphics[width=16.5cm]{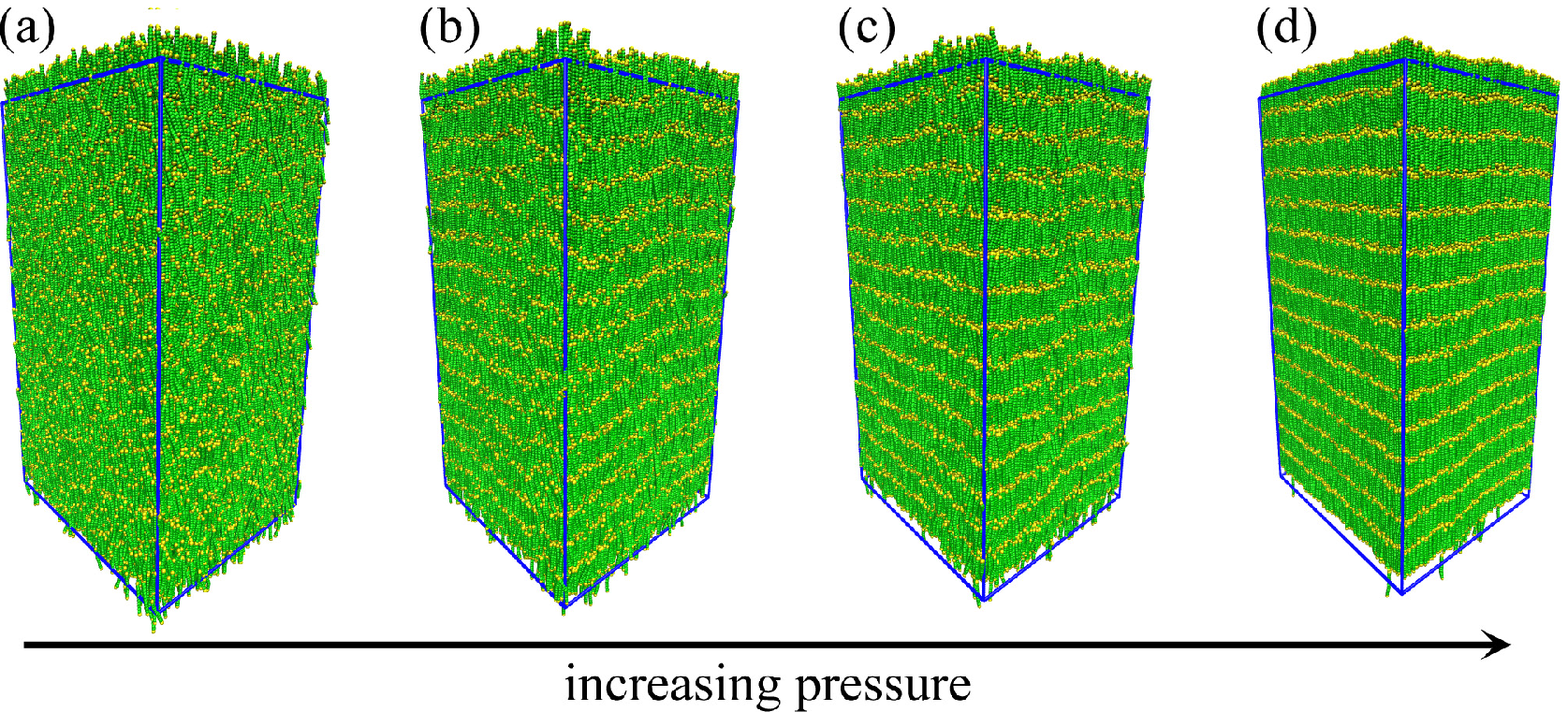}
	\caption{Snapshots of a system with ${\cal N}=89600$ chains of length $N=12$
	at $k_{\rm B}T=0.5$ for $\kappa=64$ and four different pressures
	(a) $P=0.6$, (b) $P=0.7$, (c) $P=0.8$, and (d) $P=0.9$, as indicated. Case (a)
	shows a resulting nematic state, cases (b), (c) are typical smectic states,
	while case (d) refers to a somewhat disordered crystal. Monomers are shown in
	green, apart from the chain ends which are shown in yellow and the edges of
	the simulation box are indicated by straight blue lines. In order to reduce
	finite size effects, the linear dimension of the simulation box in the
	$z$-direction was chosen about twice as large as in the $x$, $y$-directions.}
	\label{fig_3}
\end{figure*}

Fig.~\ref{fig_3}, as a preview of results whose precise analysis will follow
below, shows typical snapshot pictures of chain with $N=12$, $\kappa =64$ at
$k_{\rm B}T=0.5$ for the (a) nematic, (b, c) smectic, and (d) crystalline phases
of this model, as obtained in the ${\cal N}PT$ ensemble when all linear
dimensions $L_x$, $L_y$, and $L_z$ are allowed to fluctuate independently.

\section{Results}
\label{sec:results}
\subsection{Phase diagrams and chain center of mass distribution functions}
\begin{figure}[htbp]
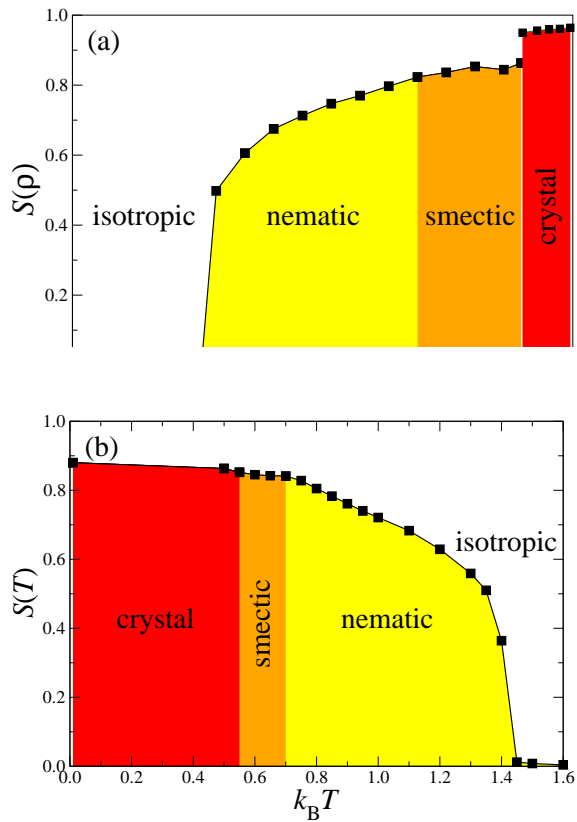

	\includegraphics[width=7.5cm]{{figures/OP_phi_N8K16T0.5}.eps}
	\quad
	\includegraphics[width=7.5cm]{{figures/OP_T_N8K16phi0.8125}.eps}
	\caption{(a) Nematic order parameter, $S$, vs density, $\rho$, in a system
	with $N=8$,	$\kappa = 16$, $k_{\rm B}T=0.5$ and ${\cal N}=101568$. Simulations
	have been conducted	in the ${\cal N}VT$ ensemble, using a simulation box with
	$L_x = L_y$, and $L_z = 100$. The lateral linear dimensions, $L_x$ and $L_y$,
	were adjusted such that the shown values of $\rho$ resulted. (b) Nematic order
	parameter, $S$,	plotted vs $k_{\rm B}T$, in the same system as panel
	(a), using a cubic simulation box with $L_x = L_y = L_z = 100$ and $\rho=0.812544$.}
	\label{fig_4}
 \end{figure}

We first focus on a system with $N=8$, $\kappa=16$, ${\cal N} = 101568$ at  
$k_{\rm B}T=0.5$. While previous work on the isotropic-nematic
transition \cite{SEAMKB,SEAMPVKB,AMSEKBAN} did always choose $N \geq 16$ (and 
$k_{\rm B}T=1.0$), we deliberately study here shorter chains first, since then
finite-size effects associated with the small number of smectic layers are
expected to be less relevant. But for $N=8$ and $k_{\rm B}T = 1.0$ the
isotropic-nematic transition is shifted to a high density already so that a
possible smectic phase would be hard to distinguish from a crystalline phase,
which we expect at densities in the order of $\rho = 1.0$ (i.e. when the
monomers effectively ``touch''). Figure.~\ref{fig_4}a, shows the nematic order
parameter $S$ (see definition below) as a function of $\rho$ at $k_{\rm B}T=0.5$,
demonstrating that the phases are well separated from each other at the lower
temperature; the isotropic-nematic transition occurs at density $\rho \approx 0.59$
while the nematic-smectic transition takes place at density $\rho \approx 0.74$.
For densities $\rho \gtrsim 0.81$ the semiflexible polymers obtain hexatic
(or even crystalline) order.

The nematic order parameter $S$ is defined as the largest eigenvalue $\lambda_3$
of the tensor $Q^{\alpha \beta}$, characterizing the average bond orientational
order of the chains; denoting $\mathbf{u}_{ij}$ a unit vector along 
the bond vector $\mathbf{a}_{ij}$, referring to effective monomer $i$ of chain $j$,
we have 
\begin{equation}
	Q^{\alpha \beta} = \frac{1}{2} \left(\la 3 u_{ij}^\alpha u_{ij}^\beta 
	\ra - \delta^{\alpha \beta} \right), 
	\label{eq_6}
\end{equation}
where the average $\langle \ldots \rangle$ is both a temporal average and an 
average over all the ${\cal N}(N-1)$ bonds in the system. In general, the tensor
$Q^{\alpha \beta}$ has three eigenvalues $\lambda_3 > \lambda_2 > \lambda_1$, but
in the nematic phase the biaxiality $B = (\lambda_2 - \lambda_1)/2$ is zero, and
since $Q^{\alpha \beta}$ is traceless, we must then have $\lambda_2 = \lambda_1 =
-S/2$. We have computed $B$ as a check, and find indeed $B \approx 0$ in the
nematic phase (within statistical error). Slightly nonzero values of $B$ occur in
the smectic phase, however. This slight increase of $B$ may indicate minor
banana-shape distortion of the chains, resulting from a misfit of the smectic
layer in the simulation box. This preliminary identification of the nature of the
smectic phase (as well as the observation of chain distortion) is suggested by
snapshot pictures, similar to Figs.~\ref{fig_1},\ref{fig_3}.

An alternative description of the global phase diagram is possible by retaining
a density $\rho = 0.812544$ and varying the temperature, $T$, as shown in Fig.
\ref{fig_4}b. It is found that the distortion of the smectic phase (measured by
a small but definitely nonzero biaxiality $B$) persists up to the transition to
the nematic phase at about $k_{\rm B}T = 0.69$. This transition does not involve
a discontinuity in $S$, suggesting (as Fig.~\ref{fig_4}a does) that the
nematic-smectic A transition is continuous.

\begin{figure}[htb] 
	\includegraphics[width=7.5cm]{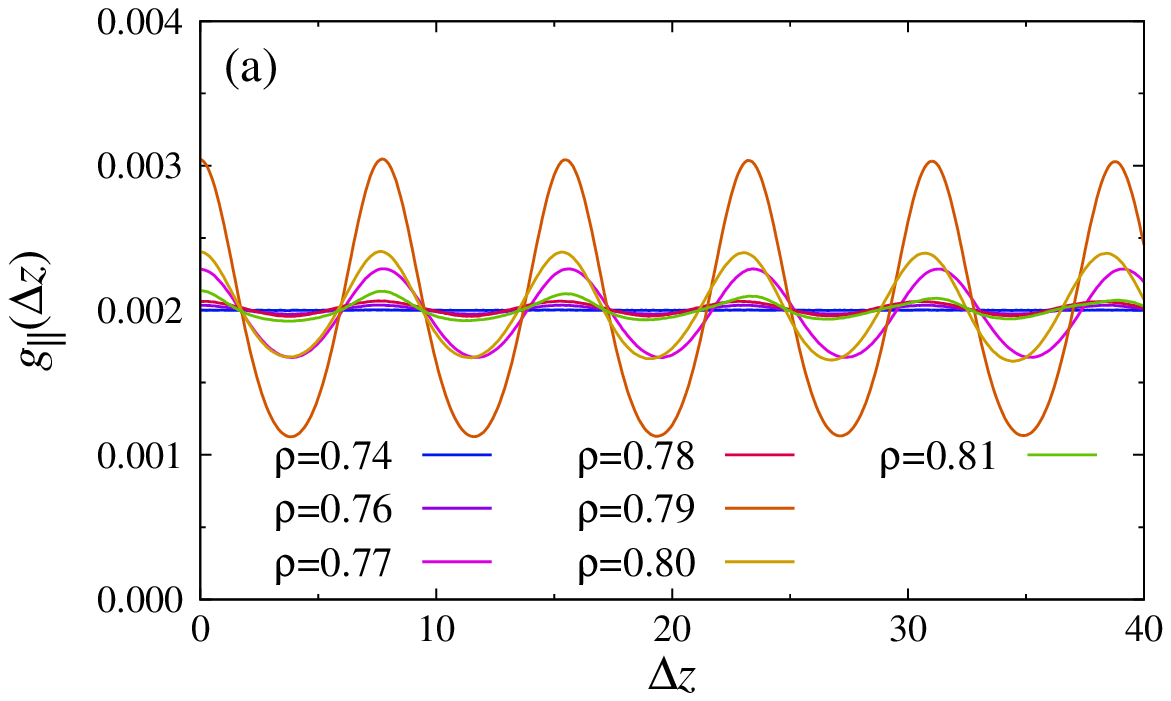}
	\quad
	\includegraphics[width=7.5cm]{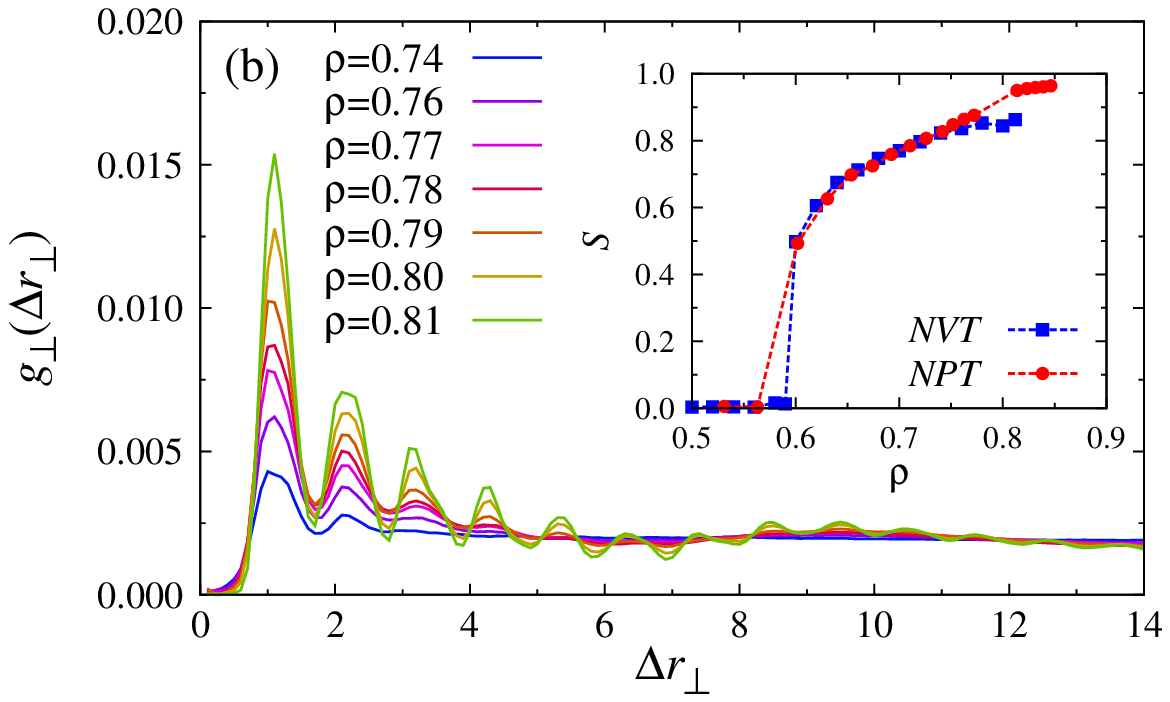}
	\caption{(a) Distribution function $g_\parallel(\Delta z)$ of the distances of
	the	centers of mass of the chains in $z$-direction for the case $N=8$, $\kappa=16$,
	${\cal N}=101568$, $k_{\rm B}T=0.5$, plotted vs $\Delta z$ for 7 densities. Data
	from ${\cal N}VT$ simulations in a cubic simulation box. (b) Radial distribution
	function $g_\perp$ of the chains center of mass positions for the same system as
	in (a). The inset shows the corresponding variation of $S$ with density in
	the ${\cal N}VT$ and ${\cal N}PT$ ensemble.}
	\label{fig_5}
\end{figure}

\begin{figure}[htb]
	\includegraphics[width=7.5cm]{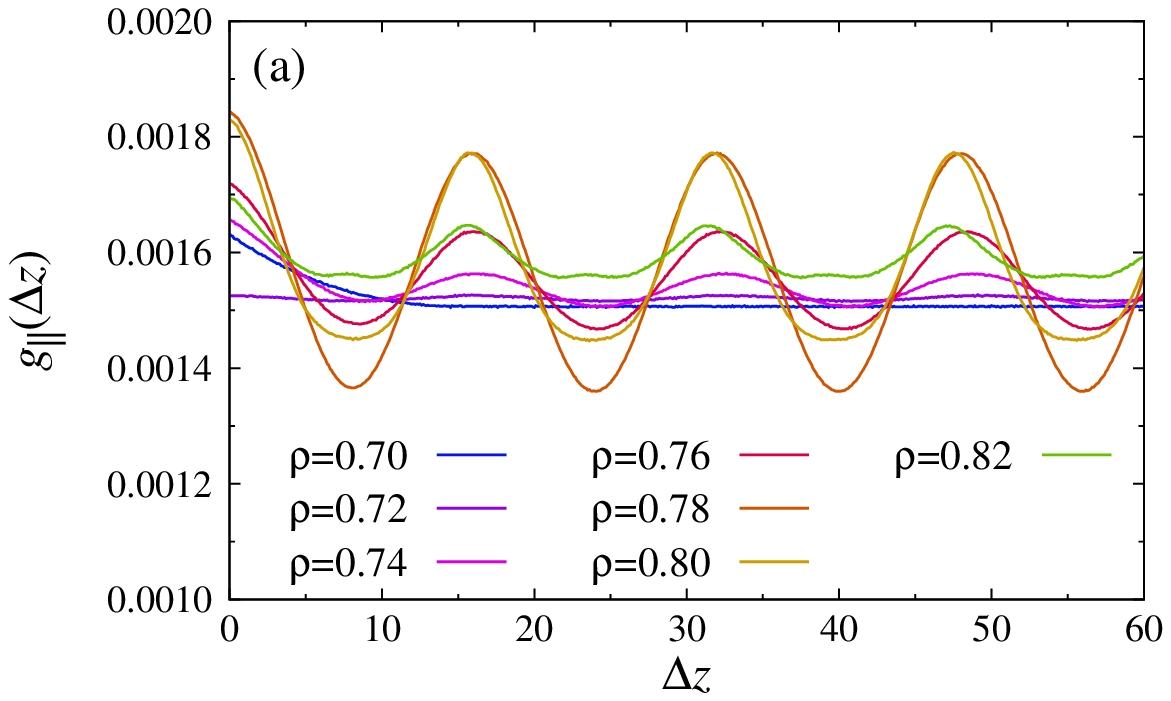}
	\quad
	\includegraphics[width=7.5cm]{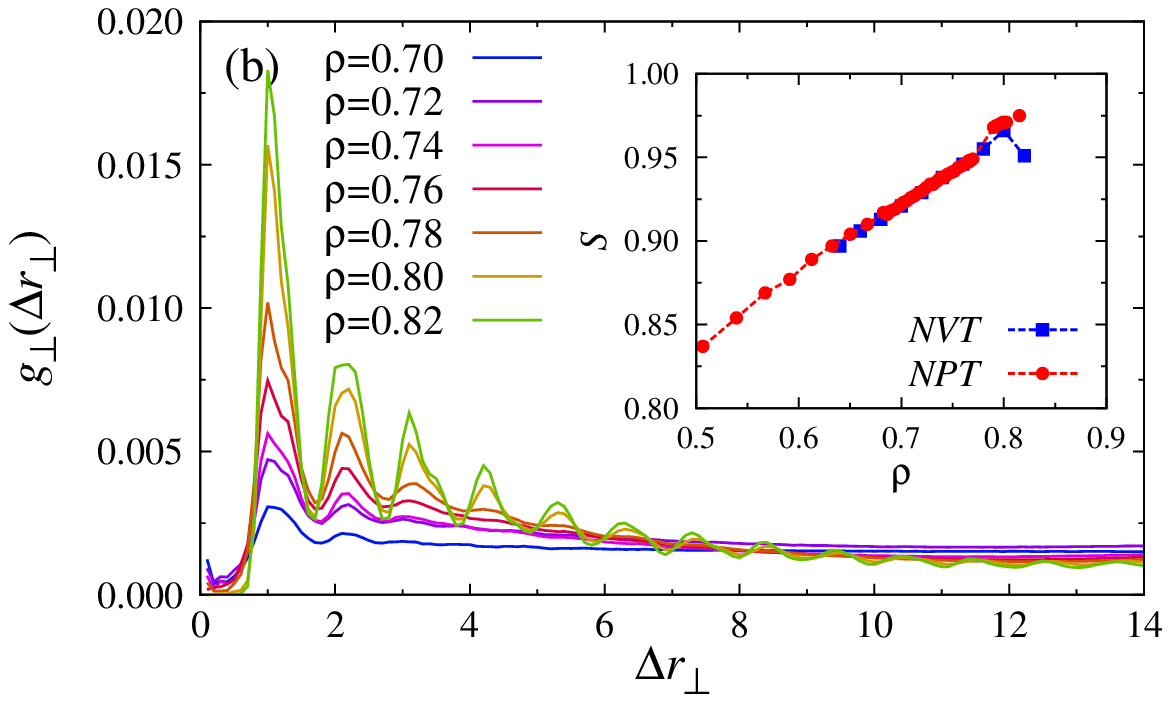}
	\caption{(a) Distribution function $g_\parallel(\Delta z)$ of the distances
	of the centers of mass in the $z$-direction for the case $N=16$, $\kappa=64$,
	${\cal N}=101568$, $k_{\rm B}T=1.0$, plotted vs $\Delta z$ for 7 densities. Data
	from ${\cal N}VT$ simulations in a cubic simulation box. (b) Radial distribution
	function $g_\perp$  of the chain center of mass positions for the same system 
	as in (a). The inset shows the corresponding variation of $S$ with density in
	the ${\cal N}VT$ and ${\cal N}PT$ ensemble.}
	 \label{fig_6}
\end{figure}

In order to identify smectic phases more precisely, we have studied the correlation 
functions between the center of mass positions of the chains, both along the 
$z$-direction, $g_\parallel(\Delta z)$, and in the radial direction perpendicular
to the $z$-axis, $g_\perp(\Delta r_\perp)$. Here, we denote the components of the
distance vector between the center of mass positions of two chains as $(\Delta x,
\Delta y, \Delta z)$ with $\Delta r_\perp = \sqrt{(\Delta x)^2 + (\Delta y)^2}$.
Figures~\ref{fig_5} and \ref{fig_6} show the
data for two selected cases, revealing that the onset of smectic order shows up
by means of periodic modulation of $g_\parallel(\Delta z)$. For the case of $N=8$,
$\kappa=16$, $k_{\rm B}T=0.5$ (Fig.~\ref{fig_5}), the modulation sets in at
$\rho \approx 0.77$ with a wave length of $\Lambda = 7.9$. This wavelength slightly
exceeds the minimum length $\sqrt{\la R_{{\rm e}, z}^2 \ra}+\sigma \approx 7.6$,
needed for a smectic structure. Note that at each chain's end $\sigma/2$ has to be
added to account for the excluded volume of the end monomers, and $\sqrt{\la R_{{\rm e}, z}^2 \ra}
\approx 6.6 < L \approx 6.8$ due to the slight tilt of the chains in the smectic
order. Thus there is some extra free volume gained for the chain ends in the
smectic structure, and the resulting entropy gain is in fact responsible for
stabilizing smectic rather than nematic order \cite{Tkach3, Cinacci}.

However, while for $\rho = 0.76$ the box linear dimension is about $L_x = L_y = 
L_z = 102.25$, and hence 13 periods do fit into the box, for $\rho = 0.80$ the 
linear dimension is only about $100.52$, implying a significant distortion of 
the smectic layering with the ``natural'' period $\Lambda$ as far as neither 12 
nor 13 periods would fit nicely into the box. The same conclusion emerges from 
$g_\perp(\Delta r_\perp)$, Fig.~\ref{fig_5}b. A pronounced radial correlation of the 
center of mass positions does develop with density in the same range where the 
periodic modulation in the $z$-direction is present. But it is also seen that a
weak much larger periodicity is superimposed which is likely a consequence of
the elastic deformation caused by the incommensurability of the smectic layering
with the linear dimension of the box. Due to this misfit (and pressure 
anisotropies, similar to Fig.~\ref{fig_2}) there is a small systematic error in 
the value of the order parameter $S$ as a comparison with the ${\cal N}PT$ 
results shows. 

These incommensurablity effects are even more pronounced for longer chains where
less smectic layers fit in a simulation box of similar size. In any case,
Fig.~\ref{fig_5} suggests a kind of long-range order in the direction of the
smectic modulation (the $z$-axis) but short-range order in the perpendicular
direction, as expected for a smectic which is still fluid. 

In the second example ($N=16$, $\kappa=64$, $k_{\rm B}T=1.0$), the onset of smectic order
is  clearly recognized for $\rho=0.74$, Fig.~\ref{fig_6}a, where a periodicity of 
$g_\parallel(\Delta z)$ with $\Lambda \approx 16$ is apparent. This periodicity 
becomes more pronounced for $\rho = 0.76$ and $\rho = 0.78$, as is obvious from 
the increase of the amplitude of the periodic variation. However, for $\rho \geq 
0.8$ the amplitude starts to decrease again and the comparison with the ${\cal 
N}PT$ reveals for $S(\rho)$ a systematic error again. This effect is due to an
increasing misfit of smectic order with growing density: for $\rho = 0.74$ the 
linear dimensions of a cubic box are $L_x = L_y = L_z = 130.33$ while for $\rho 
= 0.8$ we have  $L_x = L_y = L_z = 126.99$. So if the smectic order has a 
``natural'' periodicity $\Lambda$ that fits well in the box with $L_z = 130.33$,
it clearly will fit less well in the box with $L_z = 126.99$.
Again the fact that the center of mass correlation in the transverse direction 
(Fig.\ref{fig_6}b) exhibits pronounced short-range order, but no long-range 
order, provides strong evidence that one deals here with a smectic yet not a 
crystalline phase (see also discussion in Sec.~\ref{sec:order} below). 

It is remarkable that despite the significant distortion of the smectic 
structure the resulting values for the order parameter $S(\rho)$ in 
Fig.~\ref{fig_5}a differ only marginally from those shown in Fig.~\ref{fig_4} 
for the $L_x = L_y = 100$ geometry. But it is clear that also in this case there 
is an inevitable misfit of the natural periodicity of the smectic structure. To 
obtain more reliable data, the choice of ${\cal N}PT$ simulations where $L_x$, 
$L_y$, and $L_z$ are allowed to fluctuate independently is indispensable.

\begin{figure*}[htb]
	\includegraphics[width=7.5cm]{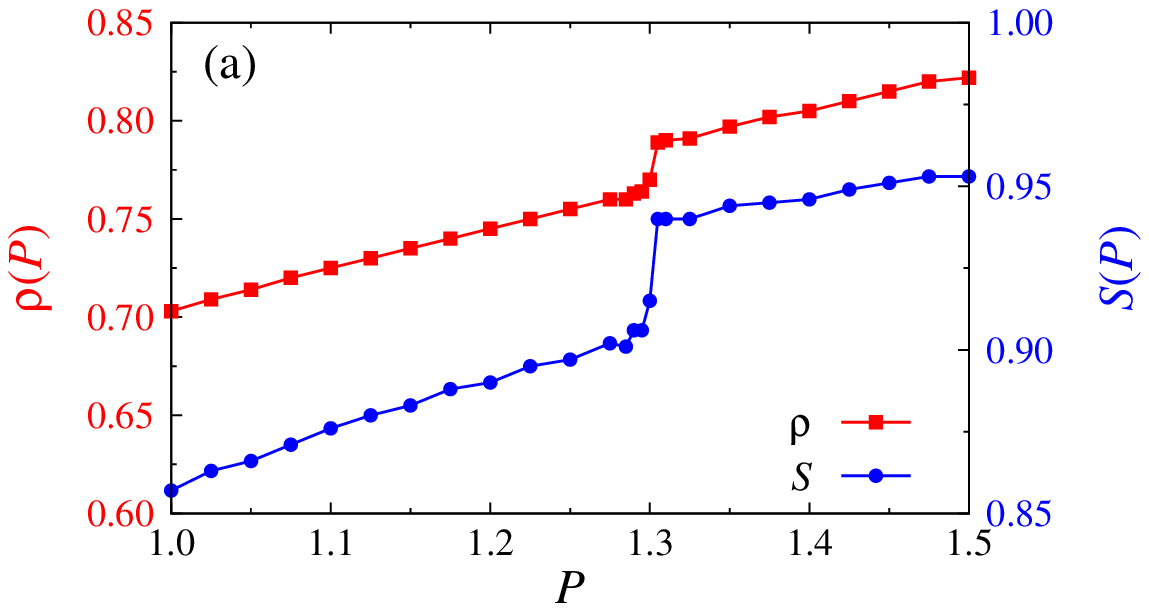}
	\quad
 	\includegraphics[width=7.5cm]{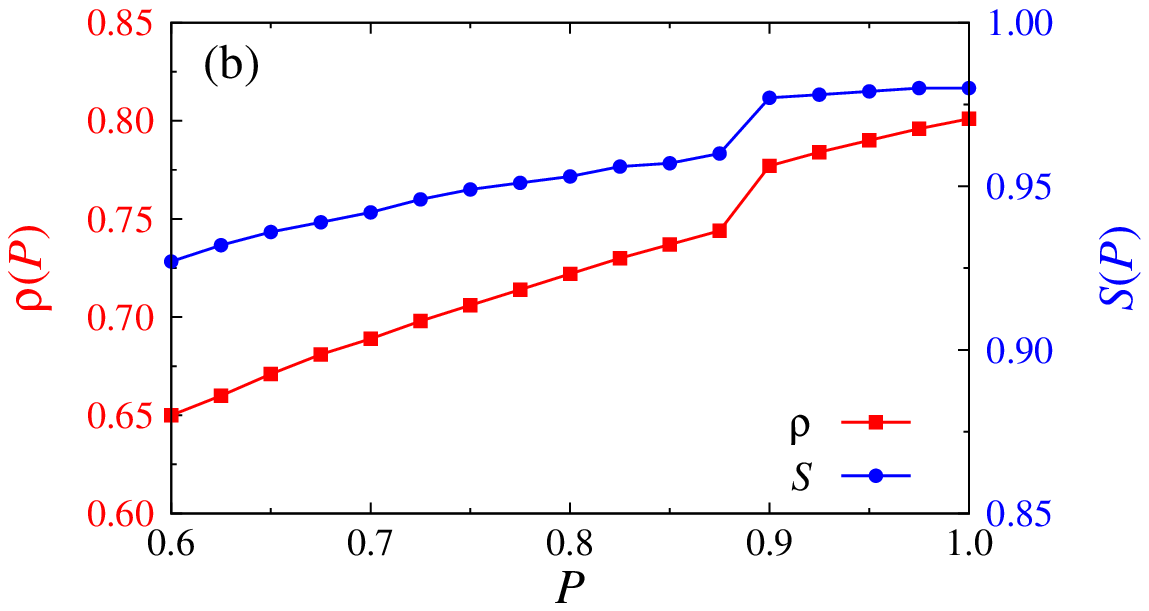}
 	\\
 	\vspace*{0.3cm}
 	\includegraphics[width=7.5cm]{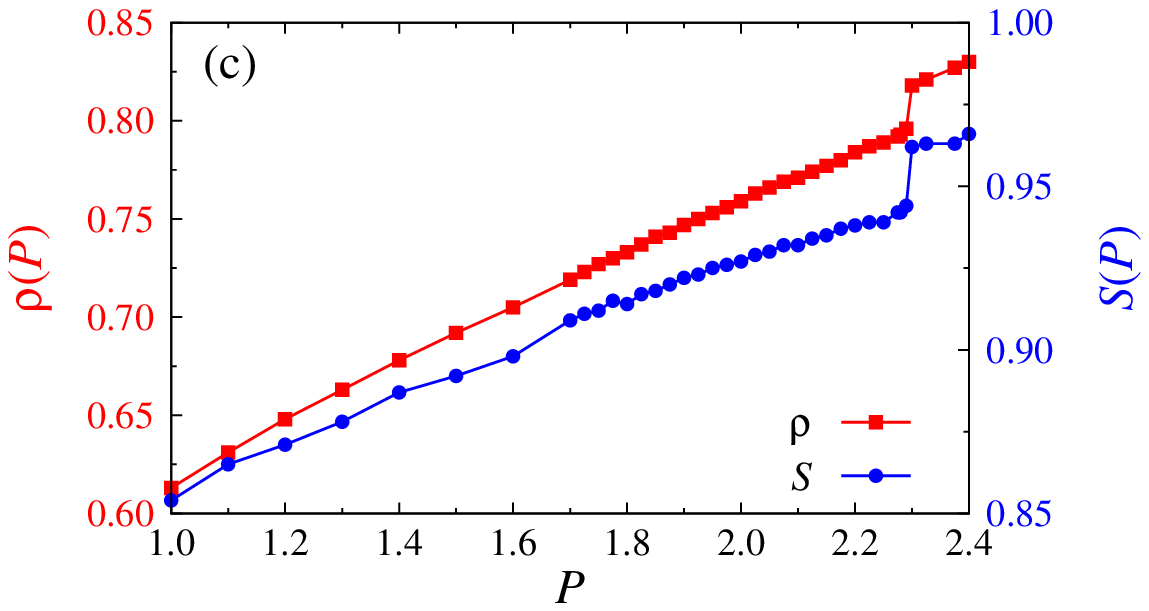}
	\quad
 	\includegraphics[width=7.5cm]{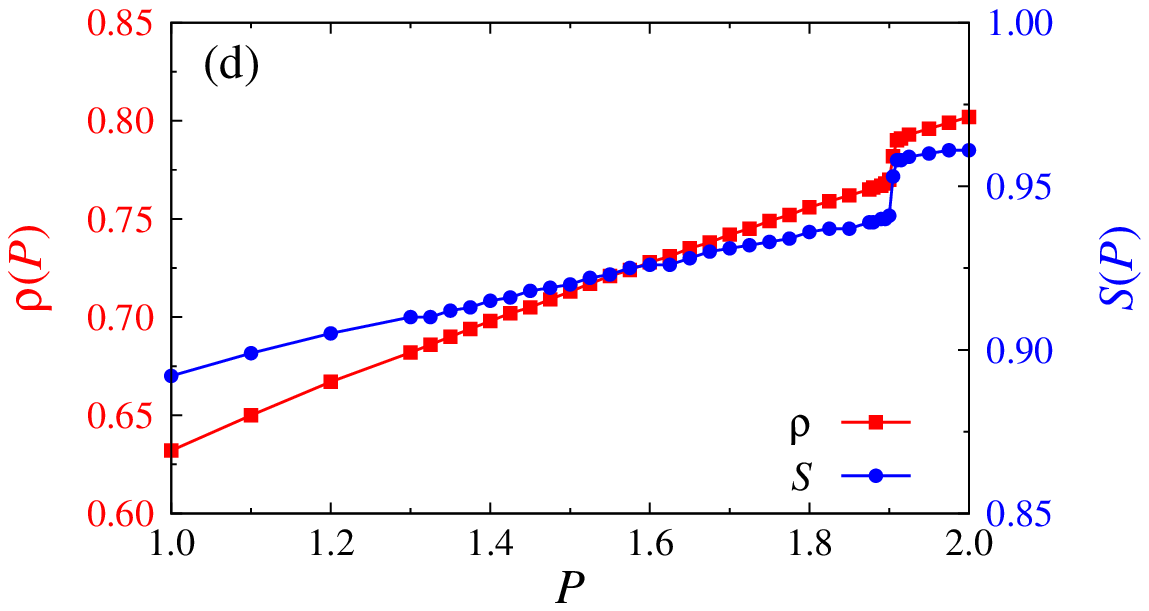}
	\caption{(a) Density, $\rho$, vs pressure, $P$, isotherms for a few
	representative cases: (a) $N=12$, $\kappa=16$, $k_{\rm B}T=0.5$, (b)
	$N=12$,	$\kappa=64$, $k_{\rm B}T=0.5$, (c) $N=12$, $\kappa=64$, 
	$k_{\rm B}T=1.0$, and (d) $N=16$, $\kappa=64$, $k_{\rm B}T=1.0$.
	The right axis in these figures show the nematic order parameter
	$S$ vs $P$.}
	\label{fig_7}
\end{figure*}
 
One of the clear advantages of the ${\cal N}PT$ ensemble emerges when we study 
the equation of state of the system: plotting the isotherm density against 
pressure, a first order transition between phases with a different character of 
the order shows up via two distinct branches separated by a density jump. In 
the ${\cal N}VT$ ensemble, the region of the density jump would correspond to a 
two-phase coexistence region, and often such regions are hard to analyze 
because of finite-size effects caused by interfaces. Indeed, Fig.~\ref{fig_7} 
reveals that such a first order transition does occur in our system at high
densities, typically in a region of densities $0.76 \le \rho \le 0.80$ (cf. 
Fig.~\ref{fig_7}). The nematic order parameter $S$ in this region then always 
exceeds $S = 0.9$ markedly, and shows also a jump (cf. Fig.~\ref{fig_7}).
From various analyses (such as those in Figs.~\ref{fig_5} and \ref{fig_6}) 
we have identified the phase at densities that are somewhat smaller than the 
density of this first order transition as smectic phases. The high density 
phase can be identified as a crystalline phase. In the next subsection, we
shall discuss order parameters that are suitable to characterize the order
both of the smectic phase and of even more ordered phases such as hexatic
liquid crystals and truly crystalline phases.

\subsection{Order parameters}
\label{sec:order}
\begin{figure}[htb]
	\includegraphics[width=7.5cm]{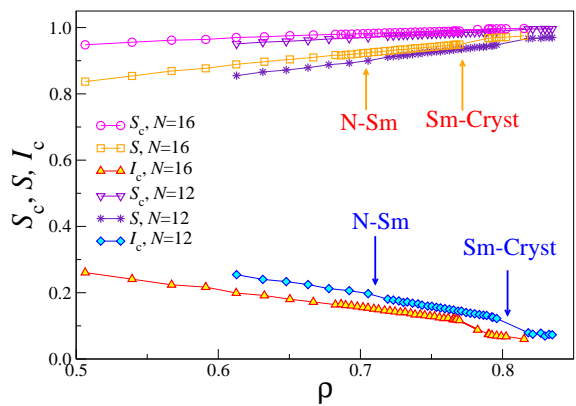}
	\caption{Chain order parameter $S_c$, Eq.~(\ref{eq_8}), bond order 
	parameter $S$, Eq.~(\ref{eq_6}), and typical chain inclination 
	$I_c \equiv \sqrt{\la R_{{\rm e}, xy}^2\ra/\la R_{\rm e}^2\ra}$ plotted vs
	density, $\rho$. Data for $N=12$ and $N=16$ for the case $k_{\rm B}T=1.0, \kappa=64$
	are included. Arrows indicate the location of the nematic-smectic transition,
	while the (first order) smectic-crystal transition in this plot shows up as
	a density gap, representing two-phase coexistence. All data were computed from 
	runs in the ${\cal N}PT$ ensemble. Note that in the smectic region $\la 
	R_{\rm e}^2 \ra$ is only marginally smaller than the results for a strictly 
	linearly stretched chain (e.g., for $N=12$, $\rho \approx 0.76$, we find 
	$\la R_{\rm e}^2 \ra/L^2 \approx 0.97$).}
	\label{fig_9}
\end{figure}

From Fig.~\ref{fig_7} it is evident that neither the density, $\rho$, nor the
nematic order parameter, $S$, show any discontinuity at the nematic-smectic 
transition. This conclusion is corroborated by a study of the nematic order of
the full chains \cite{Tortora} rather than the bonds. We define a chain order
parameter in analogy with Eq.~(\ref{eq_6}) simply in terms of the mean square
end-to-end distance components of the chains by
\begin{equation}
	S_c = \frac{3}{2}\frac{\la R_{{\rm e}, z}^2\ra}{\la R_{\rm e}^2\ra} - \frac{1}{2}
	\label{eq_8}
\end{equation}
Fig.~\ref{fig_9} compares $S_c$ and $S$, plotted vs density, for two typical 
cases. Also the typical inclination of the chains relative to the director, measured
via $I_c \equiv \sqrt{\la R_{{\rm e}, xy}^2\ra/\la R_{\rm e}^2\ra}$ with 
$\la R_{{\rm e}, xy}^2\ra = (\la R_{{\rm e}, x}^2\ra + \la R_{{\rm e}, y}^2\ra)/2$,
is shown. This inclination is typically of the order of $0.1$ to $0.25$, corresponding
to misalignments of $5^\circ$ to $15^\circ$, and it is probably due to long
wavelength collective buckling fluctuations of the nematic alignment of the chains. 

In order to characterize the smectic order quantitatively, and also locate more 
precisely the nematic-smectic phase boundary, we introduce an additional order
parameter $\tau$ that describes the periodic density modulation occurring in the
smectic phase. In an ideal smectic A phase the local monomer density along the
$z$-axis, perpendicular to the layers and coinciding with the nematic director,
is proportional to \cite{deGennes}:
\begin{equation}
	\rho(z) \propto \cos \left (\frac{2\pi}{\Lambda} z + \varphi \right),
	\label{eq_9}
\end{equation}
where $\Lambda$ is the period of the smectic modulation, and $\varphi$ is a phase 
(which is of no interest here). Eq.~(\ref{eq_9}) only holds in the smectic A
phase near the nematic-smectic transition, and neither $\Lambda$ nor $\varphi$
are known beforehand. Deeper in the smectic phase higher harmonics need to be
added to Eq.~(\ref{eq_9}). In order to determine $\Lambda$ in the general case,
it is appropriate to consider the structure factor $S(\mathbf{q})$, with wave
vector $\mathbf{q}$ oriented parallel to the $z$-axis,
\begin{equation}
	S(q_z) = \frac{1}{{\cal N}N} \la \left|\sum_j \exp(i q_z z_j)\right|^2 \ra ,
	\label{eq_10}
\end{equation}
where the sum runs over all monomers at positions $\mathbf{r}_j = (x_j, y_j, z_j)$
in the system. In the smectic phase, we expect that $ S(q_z)$ must have a rather
sharp peak at $q_z = 2\pi/\Lambda$ (see Fig.~\ref{fig:Sq} below). The smectic
order parameter, $\tau$, is then defined as the amplitude of the largest peak of
$S(q_z)$. Alternatively, one can compute the area of $S(q_z)$ underneath the
(first) Bragg peak at $q_z = 2\pi / \Lambda$, and take $\tau$ as the square root
of this area.

\begin{figure*}[htb]
	\includegraphics[width=16.5cm]{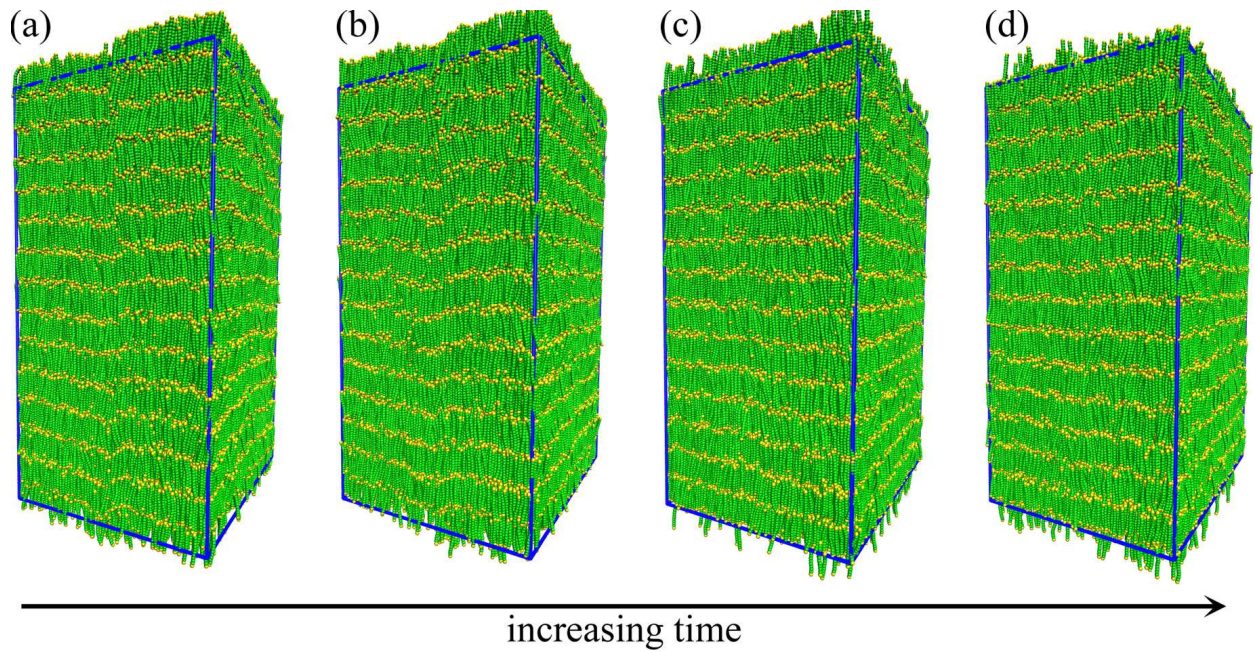}
	\caption{Snapshot pictures of the configuration of a system with $N=12$,
	${\cal N}=89600$, $\kappa=64$, $k_{\rm B}T=1.0$, and $P=2.225$ at time
	(a)	$t=1.6\times 10^4$, (b) $t=1.8\times 10^4$, (c) $t=2\times10^4$, and (d)
	$2.2\times 10^4$, as indicated. Chain ends are shown in yellow while the
	remaining monomers of the chains are shown in green. The simulation box is
	indicated by blue lines. Note that in cases (a) and (b) a spiral dislocation
	defect is present, but has disappeared in panels (c) and (d). Comparing the
	frames (c) and (d), one detects a slight difference in the phase of the
	periodic ordering, $\varphi$, Eq.~(\ref{eq_9}). Due to the random diffusive
	motion of the chains the phase is not fixed in space in the laboratory system,
	as it should be, since the smectic phase is still a fluid and not a solid.}
	\label{fig_10}
\end{figure*}

Eq.~(\ref{eq_10}) works well when one deals with relatively small systems (a few
thousand short chains were used in Refs.~\citenum{Schoot, Schoot3}). However, for
the large systems studied here (of the order of $100000$ chains) one finds often
a rather erratic behavior of the resulting order parameter when plotted vs density
or pressure. This happens because in such large systems the smectic order that
develops is nonuniform due to defects (resembling ``spiral dislocations'',
Fig.~\ref{fig_10}). Only by long annealing it is sometimes possible to heal out
these defects and obtain uniform long range order throughout the whole simulation
box, as shown in Fig.~\ref{fig_10}. To avoid the extreme investment of computer 
resources needed to achieve such annealing, we have found it more convenient to
extend the summation over the monomer coordinates in Eq.~(\ref{eq_10}) not 
over the full box, but only over subboxes $l \times l \times L_z$ with $l=5$ or 
$l=10$, respectively. We have checked that in the smectic phase the dependence 
of $\tau$ on $l$ is very weak, and that $\tau$ roughly agrees with the result 
extracted from Eqs.~(\ref{eq_10}) when uniform order is achieved in the system.
However, the drawback of this method is that in the nematic phase the resulting
$\tau$ is also nonzero. Such ``finite size tails'' of $\tau$ in the nematic phase
are familiar from similar findings at second-order phase transitions \cite{KBDWH}.

\begin{figure}[htb]
	\includegraphics[width=7.5cm]{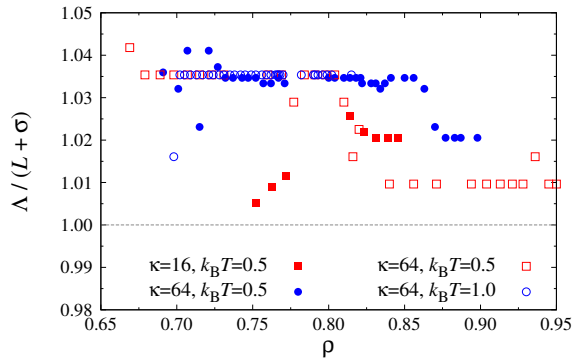}
	\caption{Period $\Lambda$ of the smectic order parameter vs density, $\rho$,
	for chains with length $N=8$ (filled symbols) and $N=16$ (open symbols),
	for various stiffnesses, $\kappa$, and temperatures, $T$, as indicated.}
	\label{fig_12}
\end{figure}

Figure~\ref{fig_12} gives a plot of the period $\Lambda$ (normalized by
$L+\sigma$) versus density for a few typical cases. It is seen that $\Lambda$ 
exceeds the estimate $L+\sigma$ slightly, indicating that the chain ends require
more space than this simple estimate suggests. The period seems to depend only
weakly on $\rho$, $T$ and $\kappa$. 

\begin{figure*}[htb]
	\includegraphics[width=7.5cm]{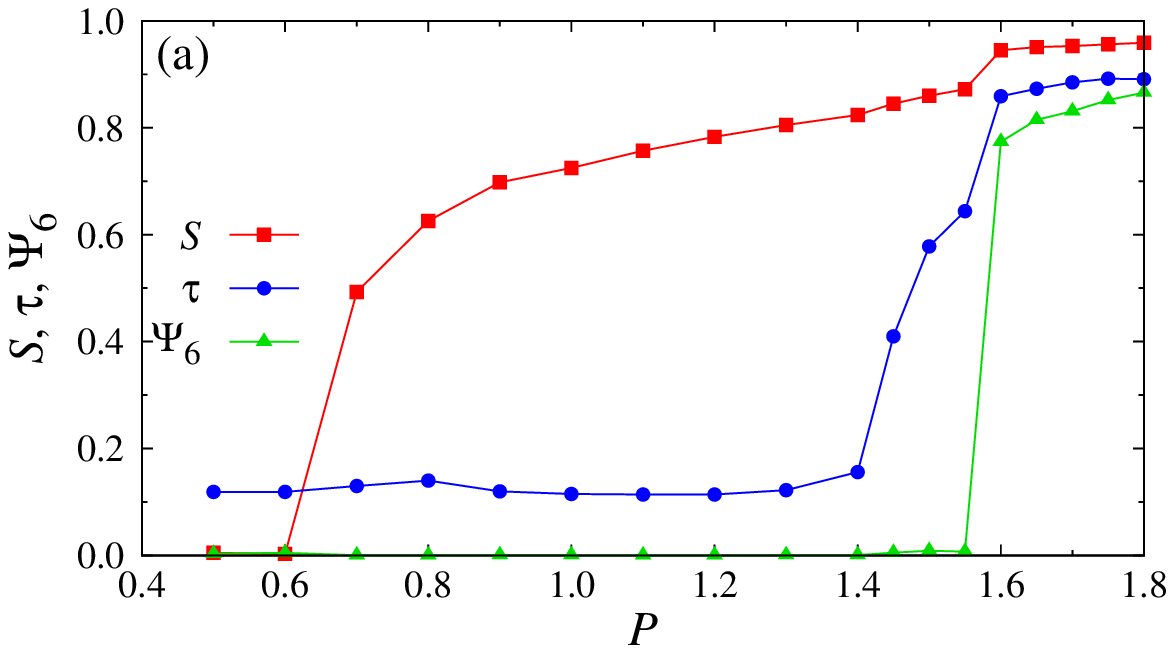}
	\quad
	\includegraphics[width=7.5cm]{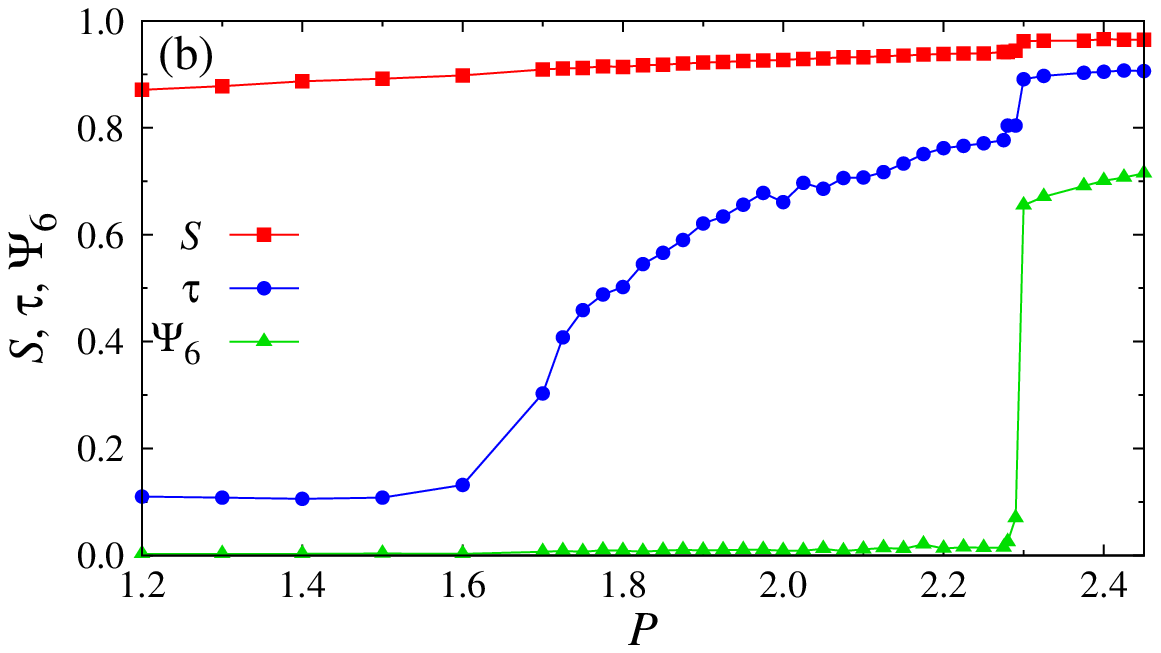}
	\\
	\vspace*{0.3cm}
	\includegraphics[width=7.5cm]{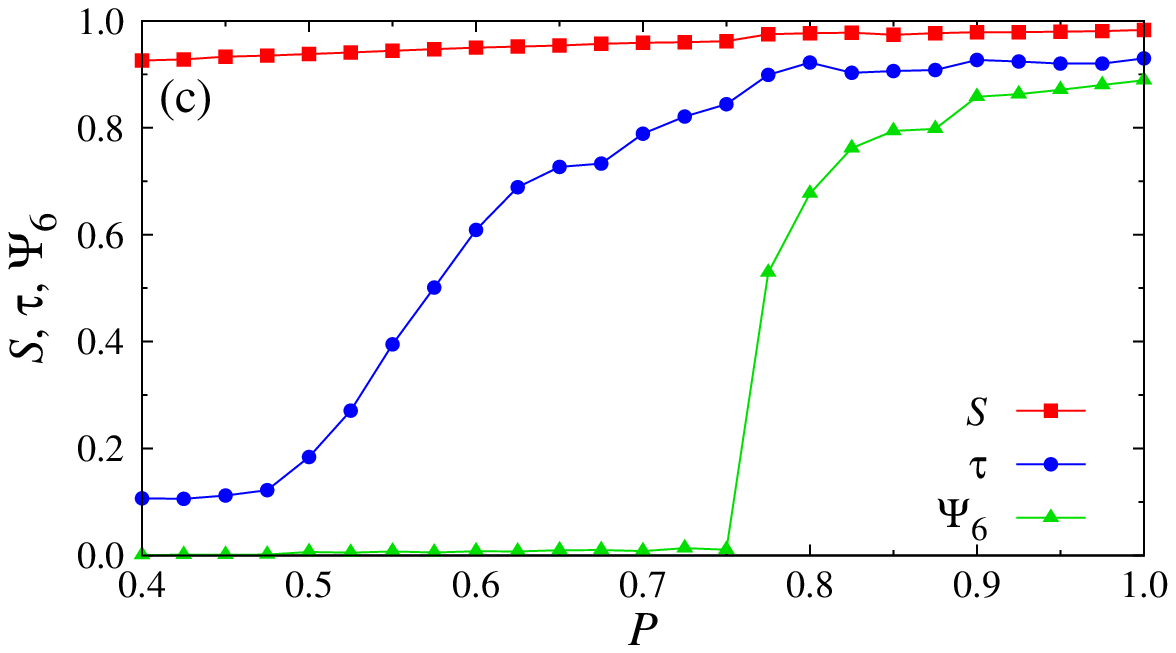}
	\quad
	\includegraphics[width=7.5cm]{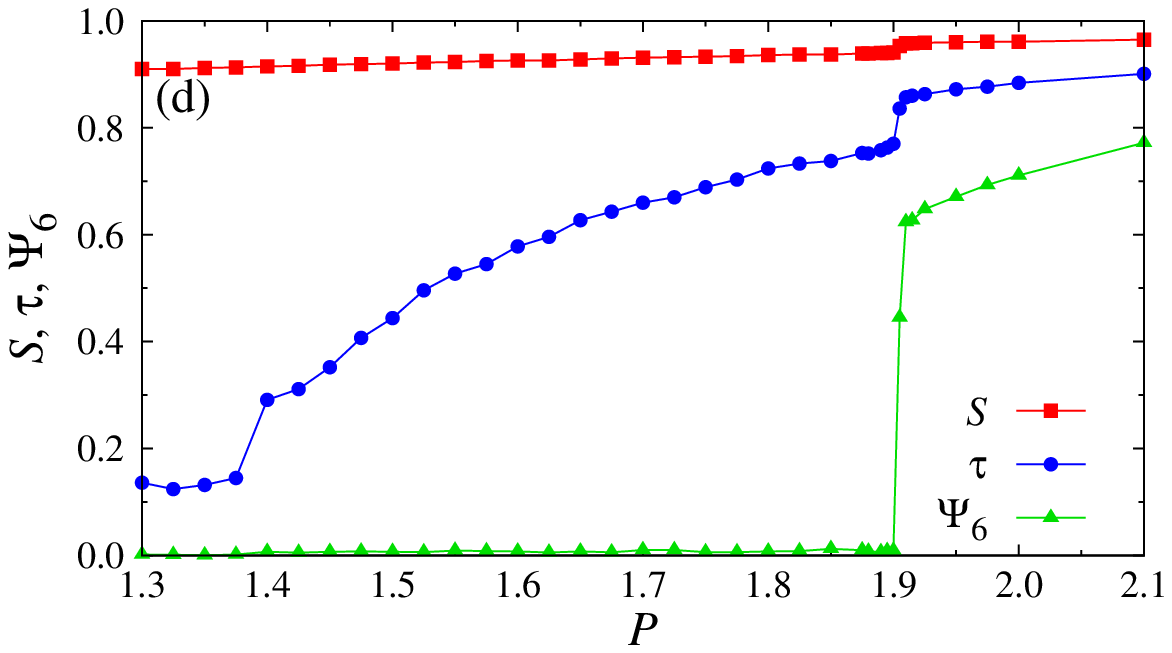}
	\caption{Variation of the order parameters $S$, $\tau$ and $\Psi_6$ with 
	pressure, $P$, for a few typical cases: (a) $N=8$, $\kappa=16$, $k_{\rm B}T=0.5$,
	(b) $N=12$, $\kappa = 64$, $k_{\rm B}T=1.0$, (c) $N=16$, $\kappa = 64$, $k_{\rm B}T=0.5$, 
	(d) $N=16$, $\kappa = 64$, $k_{\rm B}T=1.0$. Curves are intended to guide the
	eyes only.}
\label{fig_11}
\end{figure*}

Figure~\ref{fig_11} shows the typical behavior of the smectic order parameter
extracted from our simulations, revealing a seemingly continuous transition from
the nematic phase. (Because of equilibration problems and finite size effects
we did not attempt to characterize the transition more precisely.) Theoretically,
it  is well accepted that true solid-like long range order in only one dimension
is not possible \cite{Landau, deGennes2}. As these systems are at their lower
critical dimension, fluctuations prevent true long range translational order
\cite{Caille}. Also in experiments, however, the observed behavior is hardly
distinguishable from a second order transition \cite{Safinya}.

As the density is increased further, the chains need to pack more tightly and
eventually crystalline order emerges in the lateral direction. We quantify this
structuring by considering the transverse order of the center of mass 
positions $\mathbf{r}^{\rm CM} = (\mathbf{r}^{\rm CM}_{\perp, \alpha}, 
z^{\rm CM}_\alpha)$ of the chains in the individual smectic layers ($\alpha$
labels the chains in a selected smectic layer). We ask whether these coordinates 
$\{\mathbf{r}^{\rm CM}_{\perp, \alpha}\}$ form a triangular lattice order,
and therefore record the bond order parameter
\begin{equation}
	\Psi_{6\alpha} = \frac{1}{n_\beta} \sum_{\beta=1}^{n_\beta} \exp 
	(i6\phi_{\alpha \beta}),
	\label{eq_7}
\end{equation}
where the sum over $\beta$ runs over the $n_\beta$ nearest neighbors of $\alpha$
($n_\beta = 6$ in the case of a perfect lattice) and $\phi_{\alpha \beta}$ is
the angle between the vector $\mathbf{r}^{\rm CM}_{\perp, \beta} - 
\mathbf{r}^{\rm CM}_{\perp, \alpha}$ and the $x$-axis. We average 
$\Psi_{6\alpha}$ over all chains $\{\alpha\}$ in a layer and over all layers.

The $\Psi_6$ analysis can also be extended to densities in the nematic phase
where the system is arbitrarily divided into layers of thickness $L$. However,
already in the smectic phase the average $\Psi_6 = M^{-1}|\sum_\alpha \Psi_{6\alpha}|$
vanishes in the limit when the number of chains per (smectic) layer, $M$, tends
to infinity, whereas in the crystal phase the average $\Psi_6$ is clearly
nonzero. In the smectic A phase, the correlation function of $\Psi_6$ is expected
to decay exponentially with distance $r_\perp$. If a quasi-two-dimensional
hexatic phase occurs, where the lateral order of subsequent smectic layers is
decoupled, a power-law decay with $r_\perp$ is expected. However, due to
finite size effects and large statistical fluctuations, these correlations
are difficult to study, and this calculation has not been attempted here yet.

In Fig.~\ref{fig_11} we have plotted $\Psi_6$ vs pressure for selected systems.
In the smectic and nematic phase, $\Psi_6$ is essentially zero. At the transition
pressure of the crystalline phase, $\Psi_6$ abruptly jumps to values in the range
$0.7 \leq \Psi_6 \leq 0.9$ and increases with $P$ for the cases considered. The
small finite values of $\Psi_6$ that we find in the smectic phase are clearly a
fluctuation effect because we expect then a distribution $P(\Psi_6) \propto \Psi_6
\exp(-\Psi_6^2 M/2\chi_6)$ in the smectic phase with $\chi_6$ being an appropriate 
response function. The strong positional correlation between the center of mass 
positions in the transverse direction (Figs.\ref{fig_5} and \ref{fig_6}) in these 
rather dense fluid phases imply also rather strong bond-orientational 
correlations. These correlations lead to large values of $\chi_6$, reflected in 
the fluctuations seen in $\Psi_6$ in the smectic phase, but absent in the 
nematic phase.

\subsection{The scattering function $S(q_z)$ in the smectic-A phase}
We have already mentioned that the quasi-one-dimensional periodic order of
smectic layers is not a true long range order like in a crystal, since the
system is somewhat unstable against thermal fluctuations \cite{deGennes2, Caille,
Safinya, Nelson, Shalaginov}. This conclusion is drawn by analogy to the well
known problem that one-dimensional ``crystals'' are always disordered at nonzero
temperature \cite{Landau}. Here, we explore this analogy in more detail.

\begin{figure}[htbp]
	\includegraphics[width=7.5cm]{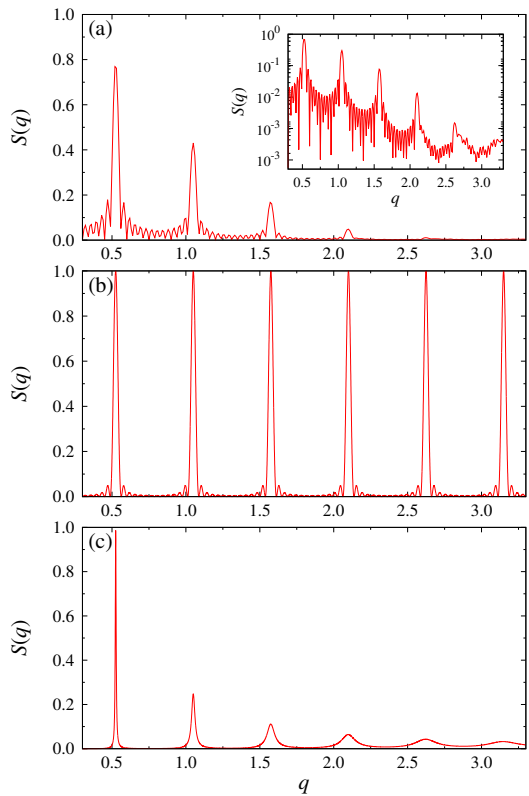}
	\caption{(a) Structure factor $S(q)$ for the system with $N=12$, $\kappa=64$,
	$k_{\rm B}T=1.0$ and pressure $P=2.225$ plotted vs $q \equiv q_z$. The inset
	shows the same data on a logarithmic ordinate scale. (b) Plot of $S(q)$ vs
	$q$, computed from Eq.~(\ref{eq:Sq}), taking $n=14$ and $\Lambda=11.97$ from
	panel (a). (c) Plot of $S(q)$ vs $q$, according to Eq.~(\ref{eq:Sq2})
	[normalized by $S(q_1)$], taking the same $\Lambda$ as in panels (a) and (b),
	and the choice $\delta^2=0.3$}	
	\label{fig:Sq}	
\end{figure}

Figure \ref{fig:Sq} shows $S(q)$ vs $q$ for a typical case. (Note that we orient
the wave vector $\mathbf{q}$ parallel to the nematic director along the $z$-axis
and omit the index $z$ from $q_z$ for simplicity). If the system were infinite
and perfectly ordered, we would expect a series of $\delta$-functions at the
Bragg positions $q_\nu = \nu 2\pi/\Lambda$, $\nu=1,2,\dots$. However, we deal
here with a finite system, which in this example consists of $n=14$ smectic
layers in total (and unlike experiments on smectic membranes where also finite
numbers of smectic layers occur, we have periodic boundary conditions rather than
free surfaces \cite{Shalaginov}). Instead of a series of $\delta$-functions at
$T=0$,  the structure factor then exhibits rather sharp peaks of finite height
(Fig.~\ref{fig:Sq}b), described by \cite{Kittel}
\begin{equation}
	S(q) = \frac{\sin\left(nq\Lambda/2\right)^2}{n^2\sin\left(q\Lambda/2\right)^2}
	\label{eq:Sq}
\end{equation}
Note that Eq.~(\ref{eq:Sq}) has minima at $nq\Lambda/2 = \mu\pi$,
$\mu=1,2\dots,n-1$, and maxima at $nq\Lambda/2 = (2\mu+1)\pi/2$, $\mu=1,2,\dots$.
Further, $S(q)$ is periodic with a period of $\Delta q = 2\pi/\Lambda$, since
$\sin\left(\pi-x\right) = \sin\left(x\right) = \sin\left(2\pi + x\right)$. The
main maxima of Eq.~(\ref{eq:Sq}) simply are $S_{\rm max}(q_\nu) = 1$, while
the heights of the side maxima near the main peaks decrease with the distance
$\Delta q = (2\mu+1)\pi/(n\Lambda)$ from the main peaks like
$4/(\Delta q \Lambda)^2 = 4/[(2\mu+1)\pi]^2$. This oscillatory ``fine
structure'' of the peaks near the Bragg positions clearly is a finite size effect,
and indeed it carries over to a large extent to the actual structure factor, Fig.
\ref{fig:Sq}a. The main difference is that the intensity of the quasi-Bragg peaks
at $q_\nu$ strongly decreases with increasing order $\nu$. This decrease of
intensity can be attributed to the effect of thermal fluctuations, which for a
one-dimensional system also would destroy long range order altogether at nonzero
temperatures. Hence, even for $n \to \infty$ the structure factor can have only
peaks of finite height and nonzero width. Assuming a harmonic one-dimensional crystal,
one obtains \cite{Emery}
\begin{equation}
	S(q) = \frac{\sinh\left(\delta^2q^2/2\right)}{\cosh\left(\delta^2q^2/2\right)
	- \cos\left(q\Lambda\right)}
	\label{eq:Sq2}
\end{equation}
where the parameter $\delta$ (with $\delta^2 \propto T$) controls the width of
the peaks. For comparison, we show Eq.~(\ref{eq:Sq2}) for $\delta^2=0.3$ in
Fig.~\ref{fig:Sq}c. We recognize a typical fluid-like structure factor, the
higher order peaks show not only a decrease in intensity with increasing order
$\nu$, but also an increasing broadening. Comparing Figs.~\ref{fig:Sq}a and
\ref{fig:Sq}c suggests that the main source of broadening for the first quasi-Bragg
peak at $q_1=2\pi/\Lambda$ are not thermal fluctuations, but finite size effects.
We have chosen here $\delta$ such that the decrease in intensity of the quasi-Bragg
peaks in Figs.~\ref{fig:Sq}a and \ref{fig:Sq}c with increasing $\nu$ is comparable
for the first few peaks. Expanding Eq.~(\ref{eq:Sq2}) for small $\delta$ and
$q \approx q_\nu$, one sees that the peak shape of $S(q)$ is Lorentzian, $S(q \approx q_\nu)
\approx \left[\delta^2 q_\nu^2/4 + (\Lambda/\delta)^2(q-q_\nu)^2/q_\nu^2\right]^{-1}$.
The peak height decreases like $S(q_\nu) \propto q_\nu^{-2}$, whereas the inset of
Fig.~\ref{fig:Sq}a would rather suggest an exponential decrease $\ln S(q_\nu) \propto -q$.
However, the finite number of layers $n$ (together with the periodic boundary
condition) prevent us from a meaningful discussion of the asymptotic behavior of
$S(q)$ for large $q$. But it is gratifying to note that the inset of Fig.~\ref{fig:Sq}a
has a remarkable similarity to corresponding specular X-ray reflectivity measurements
from smectic membranes with a comparable number of smectic layers (see, e.g., Fig. 22
of Ref. \citenum{Shalaginov}). Those measurements were done for membranes consisting
of small and rather rigid liquid crystal molecules, whose Frank elastic constants
will certainly differ from those of our lyotropic semiflexible polymers.
Experimental results for smectic phases of polymeric systems are only rarely
available, e.g. for rod-like viruses etc. \cite{Wen} and for side-group polymeric
liquid crystals \cite{Nachaliel}. The latter work observes both the first and
second quasi-Bragg peak and analyzes the shape of these peaks in terms of the
Landau-de Gennes harmonic theory \cite{Landau, deGennes, Caille}, pointing out
the significance of anharmonic effects. Such anharmonic effects may also be
relevant here, and the harmonic model [Eq.~(\ref{eq:Sq2}) and Fig.~\ref{fig:Sq}c]
should only be taken as a qualitative illustration.

\section{Conclusions}
\label{sec:conclusions}
The smectic phase of semiflexible monodisperse macromolecules in concentrated 
lyotropic solutions or melts has been investigated by molecular dynamics 
simulation of a coarse-grained bead-spring type model that was augmented by a 
bond-angle potential to account for chain stiffness. While this model is useful 
to study the isotropic and nematic phases for arbitrary ratios of the 
persistence length $\ell_{\rm p}$ and the contour length $L$, a smectic phase is 
possible only when $\ell_{\rm p} \gg L$, so that the typical macromolecular
conformation is that of a flexible rod. We have restricted our attention to
rather short chains, since the periodicity of the smectic modulation, $\Lambda$,
is close to $L$ and a large number of smectic layers must fit into the simulation
box in order to keep finite-size effects at the nematic-smectic  phase transition
at a reasonably small level.

Theory predicts that the nematic-smectic transition can be continuous. Then,
in the nematic phase both the correlation lengths of smectic fluctuations 
parallel, $\xi_\parallel$, and perpendicular, $\xi_\perp$, to the director are
expected to diverge. To investigate this behavior, we have simulated systems
containing in the order of $10^5$ macromolecules, almost two orders of magnitude
larger than previous related simulation studies. Indeed, our work suggests that
the nematic-smectic transition is continuous, while a second transition at still
higher density from the smectic phase to a more ordered (presumably crystalline)
phase is found to be unambiguously of first order, Fig.~\ref{fig_7}. While
rigorous theorems have been argued to imply that in the smectic A phase there is
no perfect one-dimensional long-range order in the direction of density modulation,
our systems still are by far not large enough to yield clear evidence for this 
phenomenon. However, despite the high density of the effective monomeric units,
we observe considerable chain inclination (both of the bonds and of the whole
chains) relative to the nematic director, leading to considerable deviations from
perfect nematic order, Fig.~\ref{fig_9}. Thus, it is clear that the correlation
lengths $\xi_\parallel$ and $\xi_\perp$ of these orientational fluctuations are
large. Even in the crystalline phase, the alignment of the rod-like polymers
along the nematic director is not yet perfect.

The crystalline phase can be detected by the presence of two-dimensional hexagonal
long-range order of the center-of-mass positions of the chains in the smectic
layers perpendicular to the director, Figs.~\ref{fig_11}. In contrast, in the
smectic phase both bond orientational correlations and positional correlations
exhibit only short-range order, Figs.~\ref{fig_5}b, \ref{fig_6}b. In the
simulations, we find that smectic order is often perturbed by the presence of
topological defects, which are difficult to anneal out, Fig.~\ref{fig_10}. It
would be interesting to search for such defects also in corresponding experiments.
While the nematic-smectic A transition has been studied extensively for small
molecule systems \cite{Safinya}, we are aware only of the observation of a
smectic phase in solutions of the tobacco mosaic virus \cite{Wen}. In that case,
the smectic layer spacing was found to exceed the contour length by about $10\,\%$
although the nematic order parameter $S$ was of the order $S \approx 0.95$.
In our model, we typically find smectic order already when $S \approx 0.9$ but
$\Lambda$ exceeds the contour length also by approximately $10\%$. Such values
are expected for short chain lengths, since $\Lambda \approx L + \sigma$ and
thus $\Lambda/L = 1 + \sigma/L$. Certainly, our model is simplified in comparison
to any real material; for instance, synthetic molecules are typically rather
polydisperse, and hence smectic order should be suppressed in comparison with
nematic order. It is a challenging problem for the future how much polydispersity
could be permissible to still allow the formation of a smectic phase.

\section*{Acknowledgments}
A.M. acknowledges financial support by the German Research Foundation (DFG) under
project numbers BI 314/24-1 and BI 314/24-2. Financial support for A.N. was 
provided also by the DFG, under project numbers NI 1487/2-1 and NI 1487/4-2.
The authors gratefully acknowledge the computing time granted on the
supercomputer Mogon at Johannes Gutenberg University Mainz (hpc.uni-mainz.de).

\end{document}